\documentstyle[11pt,newpasp,twoside,epsf]{article}
\def\edcomment#1{\iffalse\marginpar{\raggedright\sl#1\/}\else\relax\fi}
\reversemarginpar

\begin{document}
\title{Stellar Collisions and Pulsar Planets}
 \author{Brad M. S. Hansen}
\affil{Hubble Fellow, Department of Astrophysical Sciences, Peyton Hall,
 Princeton University, 
Princeton, NJ, 08544, USA}

\begin{abstract}
I describe models for the formation of planetary systems surrounding the remnants
of stellar mergers and collisions. I focus primarily on models for the viscous
evolution of disks suitable for the formation of the planets surrounding the 
pulsar B1257+12. I show that the adaptation of  models for traditional protoplanetary disks which
invoke quiescent or `dead' zones are quite successful in producing disks appropriate
for the formation of the pulsar planets. I also briefly describe some even more
exotic possibilities that may arise from compact object mergers.
\end{abstract}

\section{Introduction}
The subject of planet formation resulting from stellar encounters actually
has a long 
history.\footnote{ For a proper history of theories of solar system formation, see Brush (1990).} 
 For many years a `tidal theory' of planet formation in
our own solar system held sway (Chamberlin 1901; Moulton 1905; Jeans\footnote{I've been
unable to verify this particular reference.} 1919; Jeffreys 1929)
These postulated that the planets arose as condensations from filaments stripped
off the sun or a companion during what we would now call an exchange interaction (Lyttleton 1936)\footnote{
 Lyttleton's paper would have been perfect for
this meeting!}. Eventually this theory fell in the face of angular momentum problems
(Russell 1935), the inability of such hot filaments to condense rapidly enough
to form planets (Spitzer 1939) and the presence of Deuterium in the planets which should
have been destroyed while part of the sun (Cameron 1965).

Nevertheless, recent years have given us cause to reinvestigate some of these phenomena
in the specific context of the pulsar planets. The discovery of two earth mass planets
orbiting the pulsar B1257+12 (Wolszczan \& Frail 1992; Wolszczan 1994) began a period of
frenzied speculation as to the origin of these bodies in such an unusual location (see
Podsiadlowski 1993 for a review of the scenarios). Most scenarios proposed harken back
to the spirit of the `tidal theory' of planet formation, primarily because the evolutionary
history of the central neutron star precludes a monistic origin for these particular planets.

The purpose of this paper is to review what we think we know about the pulsar planet origins and 
to place this in the broader context of this meeting, ending with some speculation about variants
on this theme to be expected from the broad range of stellar interactions under discussion here.
In section~\ref{Origins} I will review some of the theories regarding the kinds of incidents
that led to the formation of the pulsar planets. In section~\ref{Disk} I will describe an attempt
to unify some of these scenarios in terms of a single theory of disk evolution differing only
in global parameters (essentially an extension of Phinney \& Hansen 1993). Finally in section~\ref{Others}
I will discuss a few other possible planetary systems resulting from mergers.

\section{Possible Origins for Pulsar Planets}
\label{Origins}

To properly assess the viability of the various proposed scenarios for the pulsar planets, we must
review several pieces of pertinent information. The two outermost planets (the third and smallest
planet is unimportant in terms of mass and angular momentum) are located at 0.36 and 0.47 Au,
with masses of 3.4 and 2.8 $M_{\oplus}$ respectively (the true values being larger by a factor
of 1/$\sin i$, where $i$ is the inclination angle of the planet to the line of 
sight).
This then requires a minimum of $1.3 \times 10^{48} {\rm ergs \, s}$ of angular momentum. If these
earth mass planets are rocky like our own, then an origin in solar composition material
would require a mass budget $\sim 1/Z_{\odot}$ larger, $\sim 10^{-3} M_{\odot}$. 

The observed eccentricities are small, in striking contrast to some of the more
`conventional' planetary systems recently discovered. This is a problem for those who
propose that these planets are primordial (Bailes, Lyne \& Shemar 1991),
 in the sense that they formed like any other
planetary system and survived the subsequent evolution, including the supernova that 
resulted in the pulsar. The fact that the pulsar has a 300 km/s space velocity makes
such a scenario even more unlikely.

Another illuminating fact is that this is a {\em millisecond} pulsar (spin period 6.2~ms). While most pulsars
are `young', in the sense that they have resulted from a supernova that occurred within the
last $10^6-10^8$ years, millisecond pulsars are thought to be on average somewhat\footnote{
Although I will argue later that this is not the case for this particular system.}
older (see Phinney \& Kulkarni 1994 for a review).
This is believed to result from an extended period of mass transfer from a companion, in
which the accreted angular momentum revives the old, spun-down pulsar in a new incarnation.
Thus, a natural setting for the pulsar planets is origin in a circumpulsar disk which is
also accreting onto the central object and converting it into a millisecond pulsar (Memnonides
scenarios, in the parlance of Phinney \& Hansen).
There is no shortage of such scenarios and it is these on which we will now focus.

While the circumpulsar disk is a natural ingredient for both millisecond pulsar formation and
planet formation, the source of this disk is absent. 
 Thus, we must turn to some kind of
catastrophic incident to provide the end result of a neutron star surrounded by a disk, yet
with no remaining donor.
\begin{enumerate}
 \item One possibility is simply that binary mass transfer proceeds as usual but the
 donor is eventually disrupted or evaporated. Stevens, Rees \& Podsiadlowski (1992)
 propose such a scenario in which a low mass companion is initially evaporated by pulsar
 radiation (such as appears to be happening in systems like PSR~1957+20 and PSR~1744-24).
 If the mass loss is rapid enough, the star expands adiabatically and may eventually
 suffer dynamical disruption (if sufficient mass is lost from the
 system), providing the necessary disk of material. \\
 \item A variant on this (Tavani \& Brookshaw 1992; Banit, Ruderman \& Shaham 1992) 
is that material evaporated from the companion may settle
 into a circumbinary disk. The companion may then be relentlessly evaporated into oblivion,
 leaving the remnant disk to form planets. This scenario and the former have the attractive
 feature that they take a known phenomenon to a logical conclusion. On the other hand,
 the evaporation of the companion takes $>10^8$ years, based on the observed systems. The
 high velocity and small distance above the Galactic plane suggest that the B1257+12 system
 is `young' $\sim 0.6 {\rm kpc}/ 300 {\rm km.s^{-1}} \sim 2 \times 10^6$~years. Thus, the most natural
 interpretation is that the formation of the planets and the spin-up of the pulsar to a millisecond period
 occurred soon after the supernova kick. We will use this estimate hereafter rather than the pulsar
 spin-down age ($\sim 10^9$ years) as spin-down ages are notoriously unreliable for millisecond
 pulsars (e.g. Camilo, Thorsett \& Kulkarni 1994; Hansen \& Phinney 1998)\\
 \item Another possibility is if the kick received by the neutron star happens to be
 directed at a close binary companion, so that the two suffer a direct collision. The
 tidal disruption of the companion results in the required disk. While this scenario suffers
 from the fine tuning required to engineer a collision, it is the only one which naturally gives
 rise to rapid planet formation soon after the supernova event, as required above. \\
 \item A completely different possibility is one in which the pulsar planet system
 is most closely linked with double neutron star systems. In a binary containing a
 neutron star and a massive Be star, the neutron star may capture some material from
 the Be star wind, forming the required disk (Fabian \& Podsiadlowski 1991).
 The Be star eventually explodes in a supernova,
 unbinding the binary and leaving the first neutron star with the required disk.
 This scenario has the advantage of providing a lot of angular momentum but there is no
 natural method of providing a large kick for the final system, as the second supernova
 kick is applied to the companion, not the millisecond pulsar. \\
 \item Finally, we have the case of those close, compact object binary systems which 
 undergo gravitational wave induced orbital shrinkage and merger. A merger between
 two white dwarfs could form a pulsar by accretion-induced collapse and the resulting
 remnant disk could form planets (Podsiadlowski, Pringle \& Rees 1991). This scenario
 has the twin advantages of having a known source population and providing metal-rich 
 material for the formation of planets. The main problem with this scenario is the 
 uncertain physics surrounding the issue of accretion-induced collapse (e.g. Canal et al 1990; Nomoto \& Kondo 1991).
\end{enumerate}

All of these scenarios  reach a common end-point, namely the formation of a gaseous disk orbiting the neutron star.
It is from this disk that we expect the planets to form. We may thus consider the viability of the various models
above based on the mass and angular momentum they provide to make planets in the appropriate locations.

\section{Gas Disk Evolution}
\label{Disk}

In the scenarios discussed above, disks of substantial mass ($\sim 0.1 M_{\odot}$) may form,
but generally on scales $\sim 10^9-10^{11}$~cm (thereby providing angular momenta in the range
$10^{49}-10^{52}$ ergs s, depending on the scenario), around the neutron star. This is well within
the location of the planets, so the planets are formed from the part of the disk that expands
to conserve angular momentum while most of the material is accreted (Lynden-Bell \& Pringle 1974).
Furthermore, we assume these bodies of earth mass are rocky in nature, as even the gas giants
in our own solar system require a rocky core of $\sim 10 M_{\oplus}$ before they can capture
gaseous material (e.g. Mizuno 1980). This means that only the metallicity fraction of the material
is of interest to us, thereby increasing the amount of original mass required (except, perhaps, for
the case of disrupted white dwarfs).

The expansion of such disks under the influence of simple viscosity prescriptions has been
studied by Ruden (1993) and  Phinney \& Hansen (1993). 
While the expanding disk contains enough mass and angular momentum to make the planets, by the
time the disk becomes cool enough to reasonably form dust and rocks, the material is spread over several Au, so that the planet formation
has to be both efficient and able to accumulate mass over several Au into a few planets within 1 Au.

\subsection{Constant $\alpha$ Models}

\begin{figure}                                                                                        \plotone{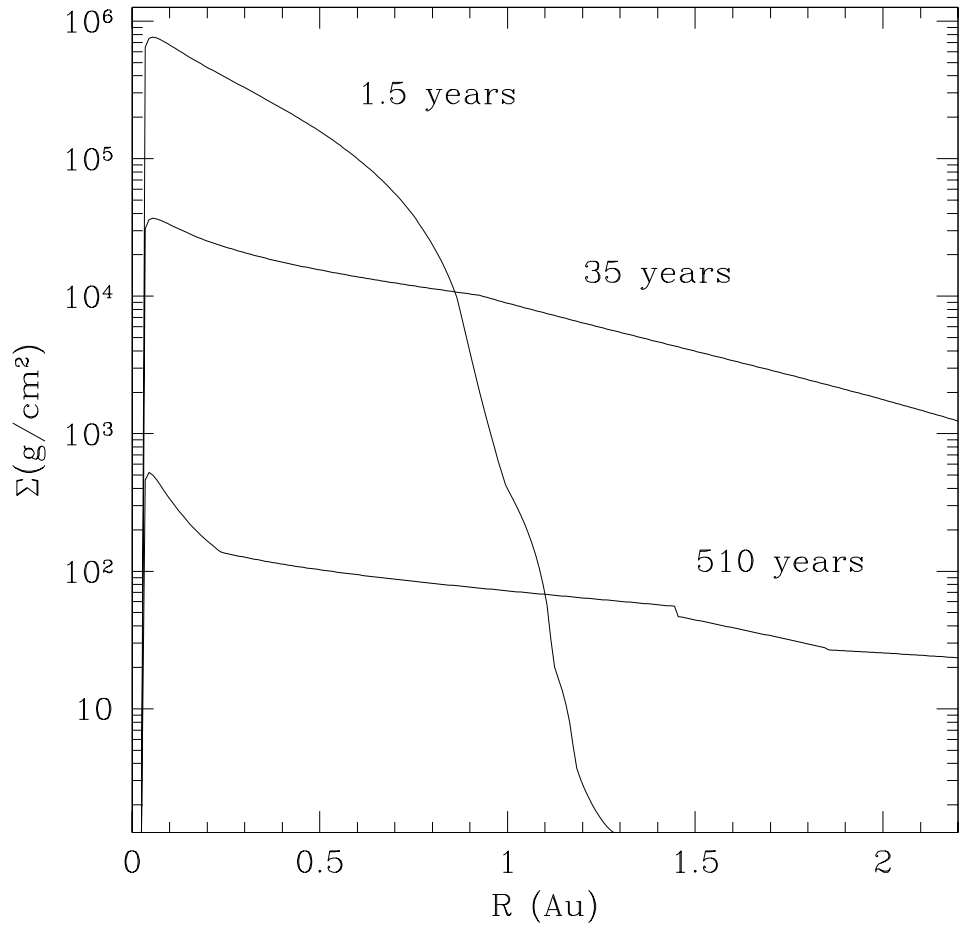}
\caption{The spreading of the disk under the assumption of a constant viscosity parameter $\alpha$ occurs quite rapidl
y, so that
most of the available planet-building mass resides at radii larger than 1 Au.
\label{Active_disk}}
\end{figure}                                                                                                                                    

To demonstrate this, consider the evolution of a disk surface density $\Sigma$ under the influence of viscosity
$\nu$, given
by mass and angular momentum conservation,
\begin{equation}
\frac{\partial \Sigma}{\partial t} = \frac{3}{R} \frac{\partial }{\partial R} \left( R^{1/2}
\frac{\partial }{\partial R} \left( \nu \Sigma R^{1/2} \right) \right).
\end{equation}

The viscosity is parameterised by an adjustable constant $\alpha$ in the standard way (Shakura \& Sunyaev 1973) such that
$\nu = \alpha c_s^2 /\Omega$, where $c_s$ is the gas sound speed in the disk and $\Omega$ is
the orbital angular frequency. Thermal balance is enforced by assuming that the energy dissipated by
$\nu$ is radiated from the surface of the disk and vertical energy transport is by radiative diffusion, modulated
by an opacity table due to Ruden \& Pollack (1991). Of particular interest to us are those low temperature regions
where the material is cool enough for dust to form. Once dust forms and agglomerates it may settle to the midplane
and eventually form planetesimals to begin the process of planetary accumulation. Thus, in this simple model
we will judge the success of the scenario by how much mass is in cool regions where material could condense in
this manner.

\begin{figure}
\plotone{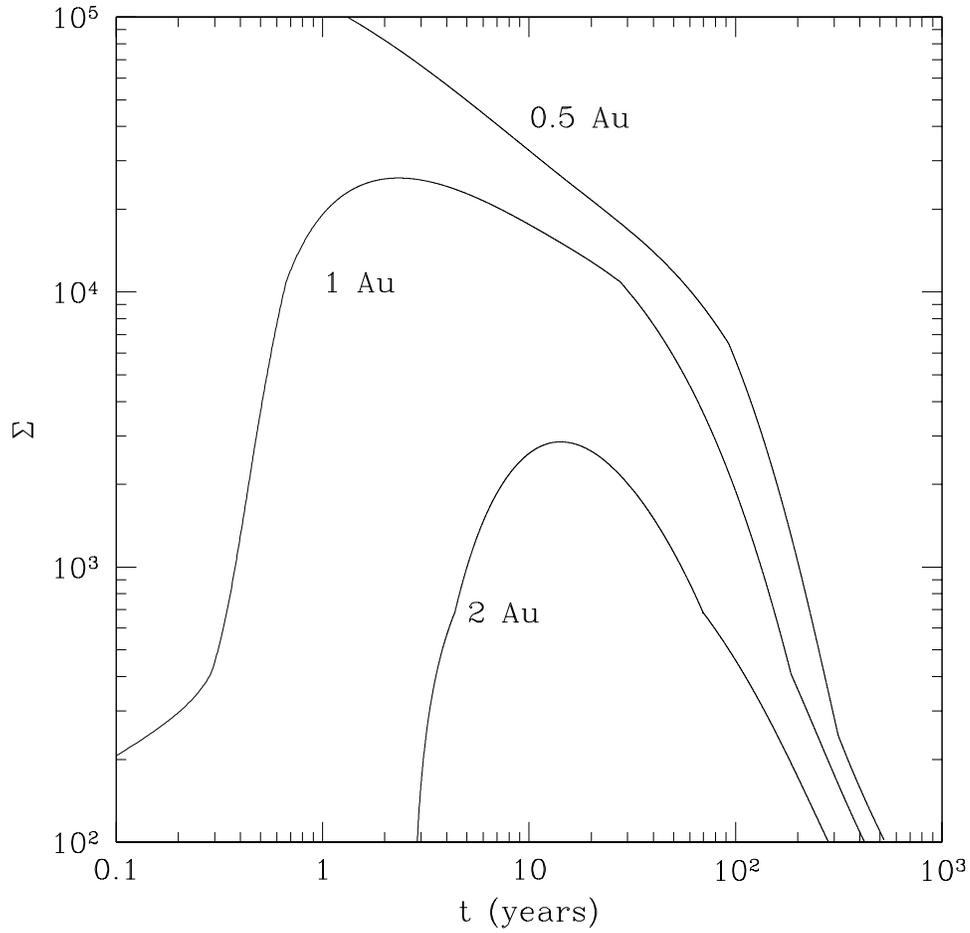}
\caption{ We see that the surface density in the interesting region peaks very early (1-10 years) and
then drops dramatically. Each curve represents the evolution of the surface density at a particular
disk radius.
\label{Radii}}
\end{figure}    

The simple $\alpha=0.1$ disk model yields uncomfortable results (Figure~\ref{Active_disk}) in that the material spreads rapidly
over several Au. Even if material could condense into solid bodies in the hotter phases, it would have to do
so rapidly, as the viscous times for the disk to expand beyond 1 Au are short,
\begin{equation}
t_{\nu} \sim \frac{R^2}{\nu} \sim \frac{6}{\alpha} {\rm years} \left( \frac{T}{3000 K} \right)^{-1} 
\left( \frac{R}{1 Au} \right)^{1/2}.
\end{equation}
 Another way to 
demonstrate this is to examine the temporal evolution of the surface density at particular disk locations (Figure~\ref{Radii}).
 Thus, any
reasonable timescale for planet formation results in material spread over many Au, much like our own solar
system. However, the planets are all located within 1 Au. This problem persists for the full range of potential
global parameters (Phinney \& Hansen 1993). Lowering $\alpha$ can prolong the period of high surface density
inside 1~Au, but still leads to a lot of mass at large radii.

\subsection{Physically Motivated Models}

\begin{figure}
\plotone{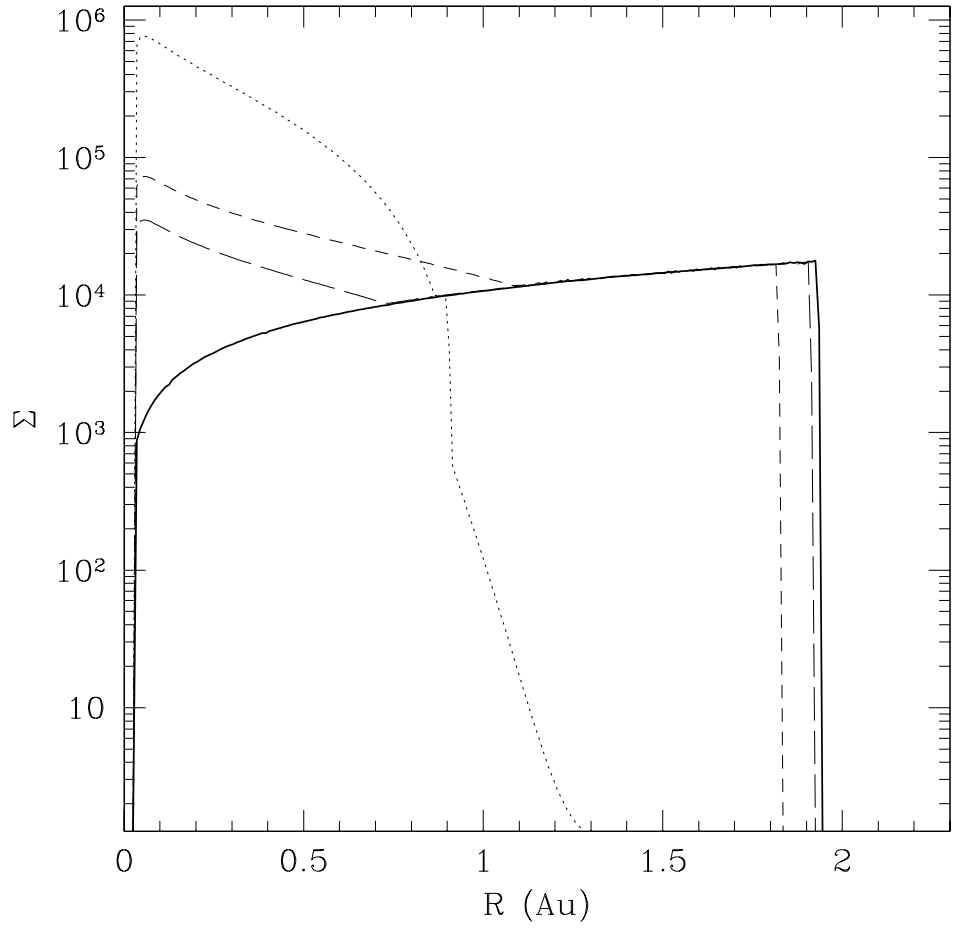}
\caption{ The solid line indicates the final asymptotic quiescent disk, while the dotted and dashed
curves indicate the evolution to the final state. \label{Absolutely_Dead}}
\end{figure}

However, there is no reason why $\alpha$ should be constant throughout the disk. This convenient assumption
is usually simply the result of having no better, physically motivated model. 
 Despite considerable work over the last few decades (see Papaloizou \& Lin 1995 for a review),
there is only one angular momentum transport mechanism that is generally regarded as robust, namely the
magnetorotational instability (Velikhov 1959; Chandrasekhar 1960; Balbus \& Hawley 1991). This mechanism
works only as long as the disk material is sufficiently ionized to couple magnetic field and gas motions.
Thus, in the outer regions of protoplanetary disks, the instability breaks down (Gammie 1996;
Sano \& Miyama 1999; Wardle 1999). If there is no other angular momentum transport mechanism at work,
then this material may be quiescent, except perhaps for a thin surface layer which is maintained at a sufficient
level of ionization by cosmic rays (Gammie 1996) or illumination from the central star (Glassgold, Najita \& Igea 1997).
Thus, as an alternative model, we may consider the expansion of a disk using the same model as before,
but where we set $\alpha=0$ when $T < 3000$~K. The end result is that the disk tends to an asymptotic 
quiescent disk, whose profile is dictated by the transition temperature and whose extent is determined by
the requirements of global angular momentum conservation. The profile corresponding to this transition
temperature is
\begin{equation}
 \Sigma = 1.1 \times 10^4 {\rm g.cm^{-2}} \left( \frac{\alpha}{0.1} \right)^{-1/2} \left( \frac{R}{1 Au} \right)^{3/4}.
\end{equation}
Integrating over this profile provides a relationship between the outer radius of quiescent disk and the total
angular momentum stored
\begin{equation}
 R_{out} = 1.61 Au \left( \frac{J_{tot}}{10^{51} {\rm ergs \, s}} \right)^{4/13}.
\end{equation}
 The evolution to this state is shown in
Figure~\ref{Absolutely_Dead}. Requiring the total angular momentum in the gas disk to be $\sim 1/Z_{\odot}$
larger than the total planetary angular momentum today, we find an outer radius of 0.7~Au. The total rocky mass
in this asymptotic disk is $\sim 7 M_{\oplus}$. Thus, this hypothesis provides an excellent order-of-magnitude
agreement with the observations.

\begin{figure}
\plottwo{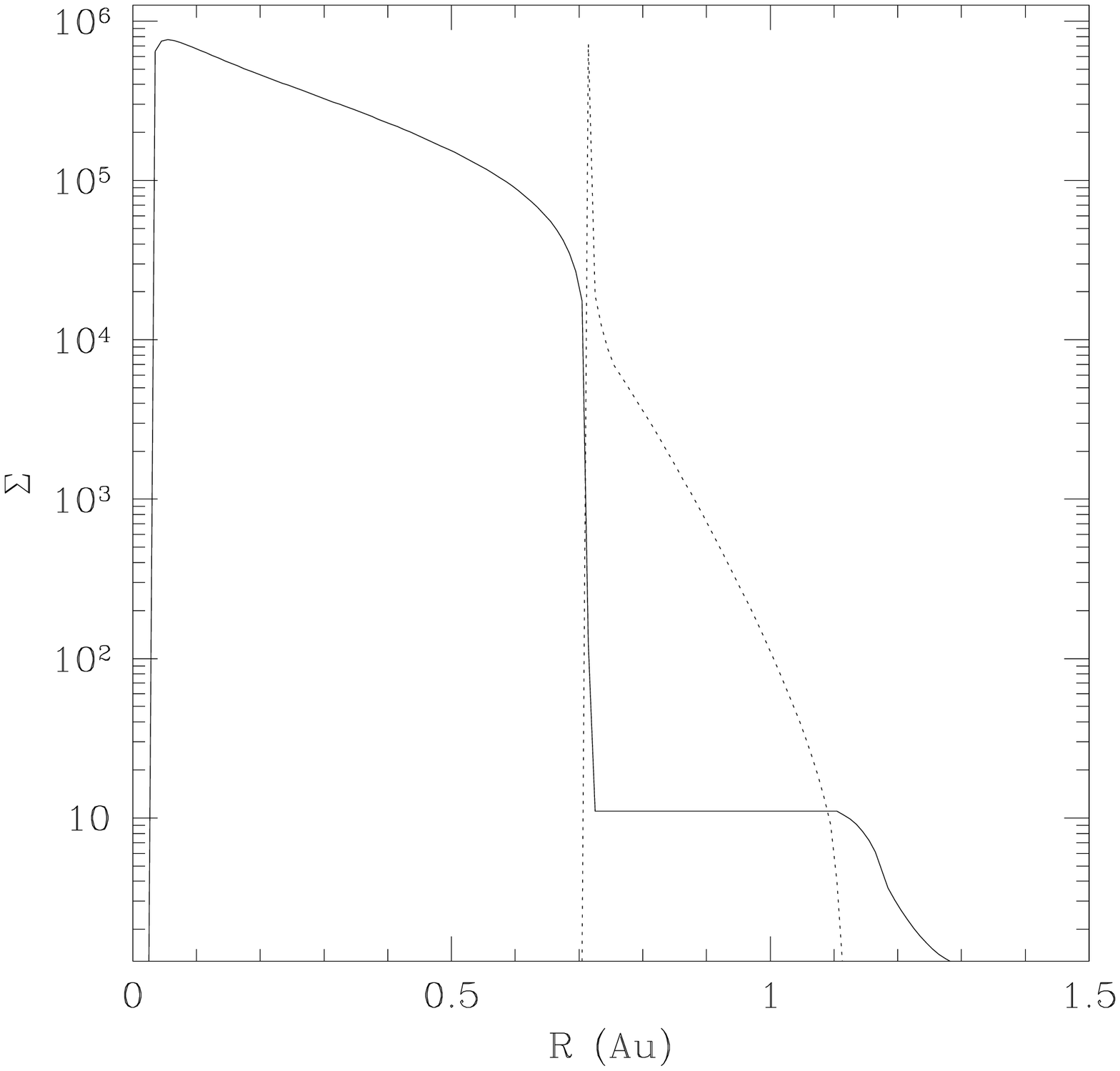}{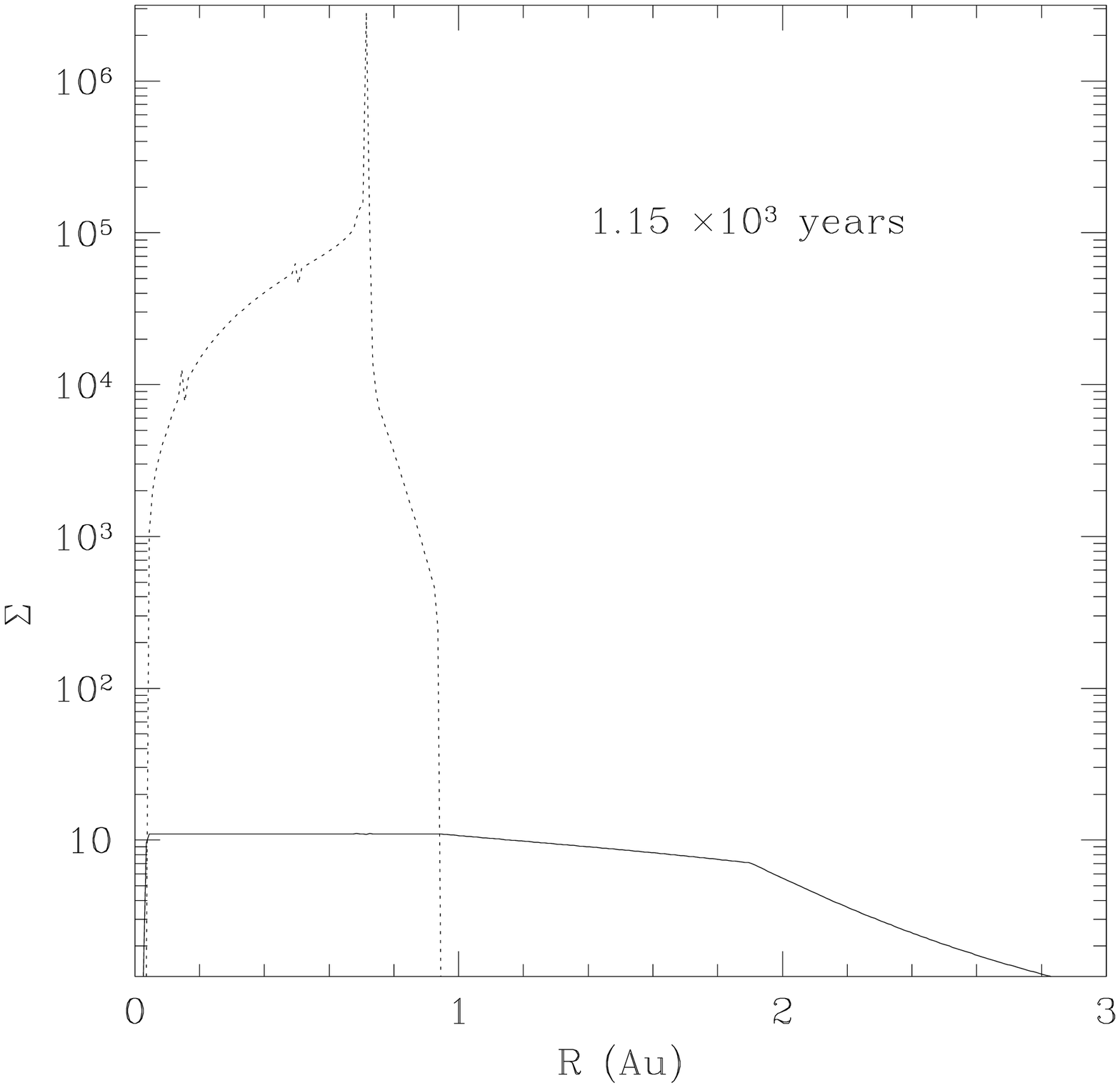}
\caption{The left hand panel indicates the early evolution of the expanding disk. The solid line
indicates the hot, active part of the disk which expands until the material is cool enough to recombine
and become quiescent (dotted line). The right hand panel indicates the end result of the evolution with
a massive quiescent disk inside 1 Au, slowly being drained by a low mass active disk maintained by
central object illumination. 
\label{A_bit_Dead}}
\end{figure}

In reality, the accretion onto the central object will result in copious X-rays, and the surface layers
of the disk may be ionized sufficiently that the magnetorotational instability may operate and some accretion
may continue. Thus, we now consider a model like the previous one, but in the spirit of Gammie (1996), in
 which a critical surface density
$\Sigma = 10 g/cm^2$ is assumed to remain sufficiently ionized to continue to evolve with $\alpha=0.1$. For
those radii where the total $\Sigma > 10 g/cm^2$, we assume the remainder is quiescent as before. This changes
the evolution somewhat, since now both the inner (by virtue of their higher temperature) and the outer
(by virtue of their lower surface density) layers remain active and the dead zone resides at intermediate
radii. It is particularly encouraging that this model produces a dense, quiescent disk at precisely the
radii where the pulsar planets reside (Figure~\ref{A_bit_Dead}).

The modifications to the original calculations now provide us with a good model, in accordance with
the broad requirements of the observed planets. We can accumulate sufficient mass and angular
momentum in cool material in the region at $\sim 1 Au$ and keep it there without spreading too
thin. The viscous timescales $\sim 10^3-10^4$ years are also short enough to be consistent with
the requirements of a young age for the system. Note however this refers only to the gaseous phase
of the evolution. The timescale requirements also demand that the heavy elements condense into
solid form, accumulate into planetesimals and thence into planets within $\sim 10^6-10^7$~years.
The formation of planetesimals is unlikely to be a problem in this regard as timescales $\sim 10^3-10^4$~years
are expected (Weidenschilling \& Cuzzi 1993).

\subsection{Accumulation Into Planets}
\begin{figure}
\plotone{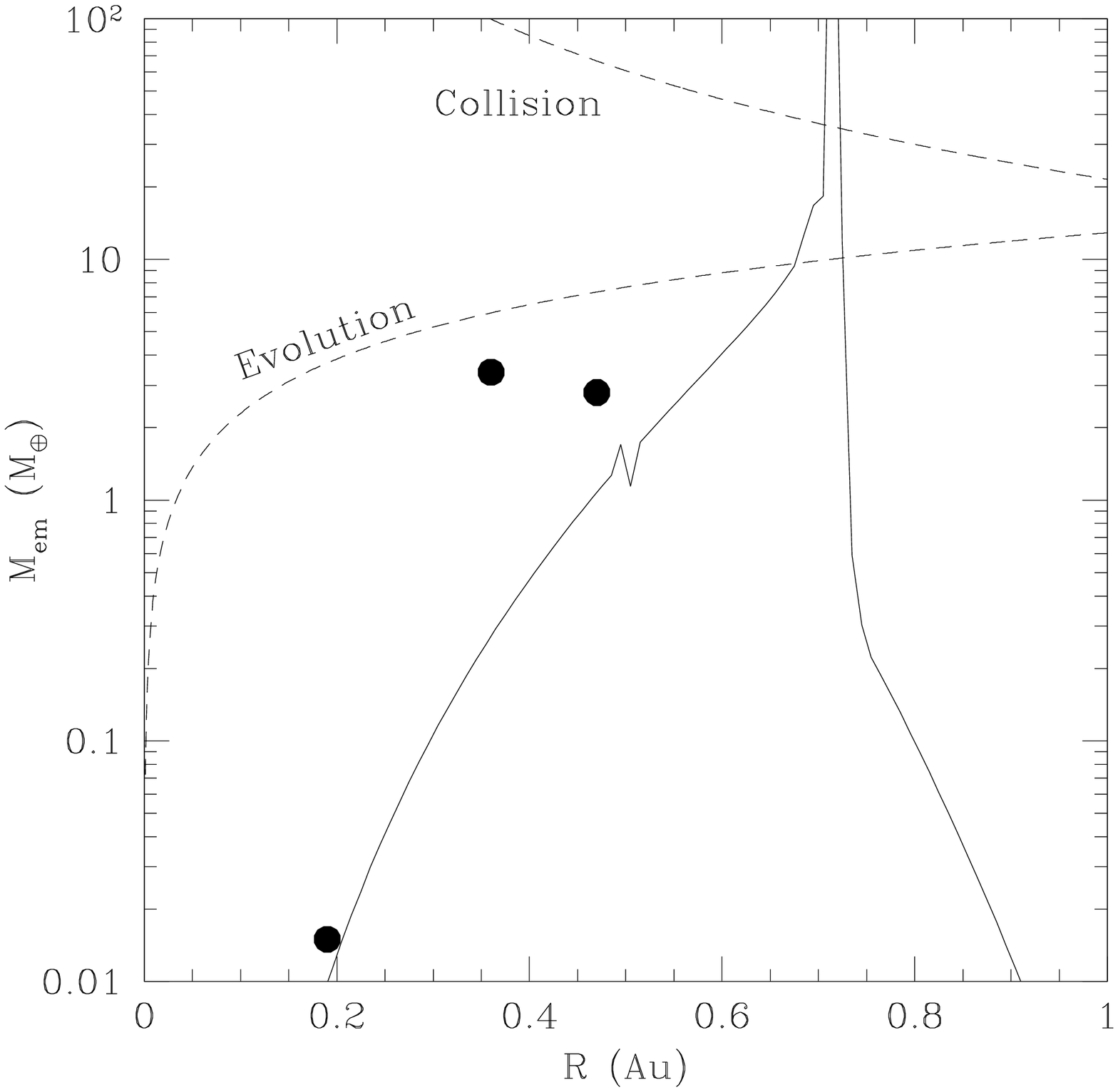}
\caption{ The solid curve is the expected embryo mass assuming the solar metallicity quiescent disk deposits
it's heavy elements into planetesimals. The three points are the low limits on the masses of the three planets.
The dashed lines labelled `Evolution' and `Collision' are described in the text.
\label{Embryo}}
\end{figure}

Assuming the formation of planetesimals occurs on timescales similar to the viscous spreading time,
we must then consider the accumulation of planetesimals into larger bodies. A rapid initial accumulation
leads to planetary embryos, isolated from each other in the sense that each has swept up the surrounding
mass in an annulus of width appropriate to the embryo's ``Hill radius'', $\Delta R \sim \mu^{1/3} R$,
where $\mu$ is the mass ratio with respect to the central object. After this, longer range gravitational
perturbations bring the isolated embryos into crossing orbits again and the accumulation proceeds.
 By taking the surface density of
the quiescent disk in Figure~\ref{A_bit_Dead}, we can calculate the expected embryo mass as a function
of radius
\begin{equation}
 M_{p} \sim 28.4 M_{\oplus} \left(\frac{R}{1 Au}\right)^{3} \left(\frac{Z}{Z_{\odot}} \right)^{3/2}
 \left( \frac{\Sigma}{10^5 g/cm^2}\right)^{3/2},
\end{equation}
which is shown in Figure~\ref{Embryo}.

We see that the innermost planet is indicative of the kind of building blocks one expects for the final 
process of planetary accumulation. The outer two planets are larger, suggesting that further accumulation
has most likely taken place. It is interesting to speculate about whether this final stage of accumulation
has, in fact, ended in this particular system, given that estimates for our own solar system indicate
this stage lasts for $\sim 10^8$ years (Wetherill 1990), i.e. somewhat longer than the likely lifetime of this system.
The final clearing stage takes place on a timescale determined by the time it takes the largest bodies
to scatter the smaller bodies onto parabolic orbits (Tremaine 1993)
\begin{equation}
t \sim 1.8 \times 10^9 years \left( \frac{M_p}{M_{\oplus}} \right)^{-2} \left( \frac{a_p}{1 Au} \right)^{3/2}
\end{equation}
The mass versus semi-major axis relation required to accomplish this feat in $\sim 10^7$~years is shown by the curve labelled
`Evolution' in Figure~\ref{Embryo}. We see that, while the known bodies are not large enough to accomplish this
feat in the expected time allotted, the largest planet is within a factor of two of the criterion, so that
significant dynamical evolution of the embryo swarm is expected. However, some relic component of the embryo
swarm may yet remain for future detection.
 It should also be noted that these bodies
are not massive enough to eject most planetesimals, rather they will accrete the smaller bodies (Tremaine 1993)
as long as
\begin{equation}
M_p < 21.5 M_{\oplus} \left( \frac{a_p}{1 Au} \right)^{-3/2} \left( \frac{\rho}{3 g /cm^3} \right)^{-1/2}
\left( \frac{\Delta i}{0.1 rad} \right)^{-3/4},
\end{equation}
where $\rho$ is the average planet density and $\Delta i$ is the average inclination of the planetesimal bodies.
This curve is also shown in Figure~\ref{Embryo}, labelled `Collision'. The planets lie well within this
bound, so that little rocky material has likely been ejected from the system (and no Oort cloud is expected).

\subsection{Further Planets In This System}

\begin{figure}
\plotone{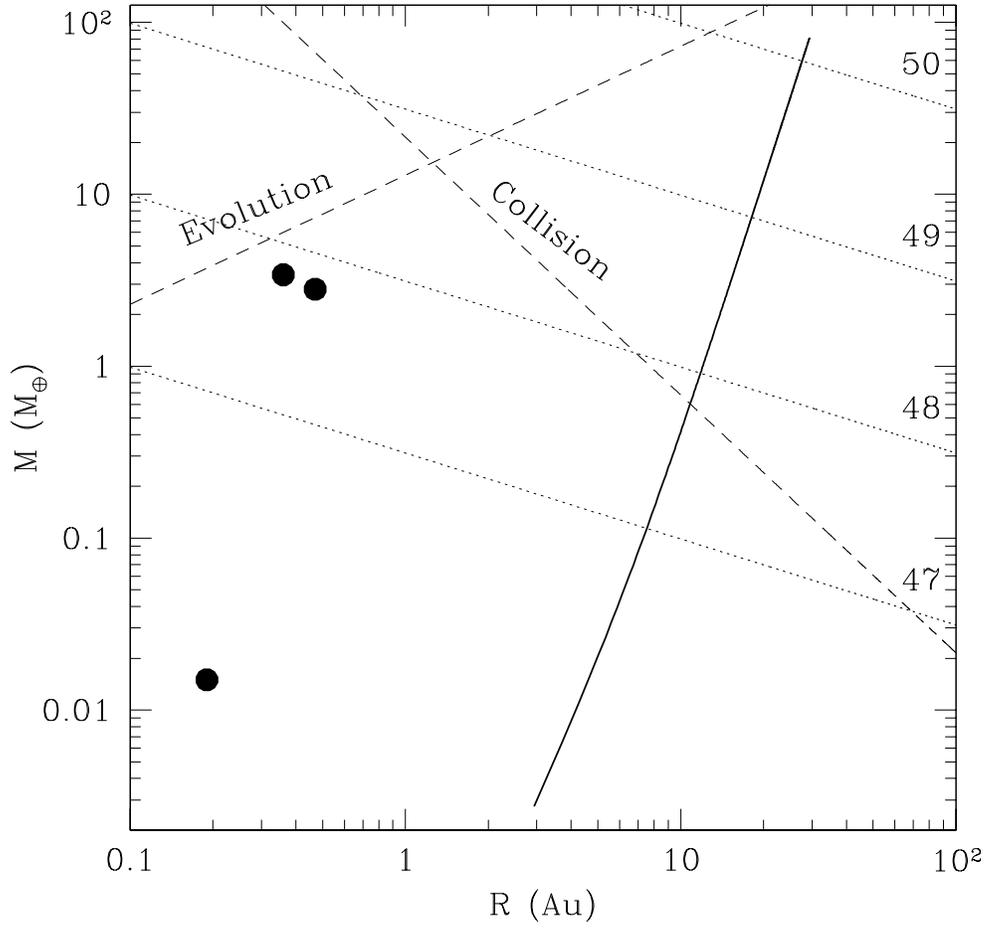}
\caption{ The solid line indicates the range of possible values for the potential fourth planet. The
dashed lines are the same as in Figure~\ref{Embryo}. The dotted lines are lines of constant angular momentum,
labelled by the logarithm of their cgs values. 
\label{New_Stuff}}
\end{figure}

The preceding discussion is based solely on the three planets reported by 1994. Further timing of this system
has revealed long term timing structure that could indicate the presence of additional planets in
this system (Wolszczan et al 2000). Figure~\ref{New_Stuff} shows the inferred planet parameters assuming a
single additional body in a circular orbit (Joshi \& Rasio 1997). There is considerable uncertainty in the
planet parameters, stemming largely from the uncertainties in how much of the contribution to the pulsar
first frequency derivative is due to the additional orbital motion. For bodies on the lower end of this
curve, little adjustment need be made in the above arguments. However, bodies in the upper end of
the allowed mass and distance range could well invalidate much of the above description. The problem is
that the angular momentum required to generate such an object is larger (once augmented by the lost
gas fraction) than many scenarios permit.

If the fourth body does turn out to be $\sim 80 M_{\oplus}$ at $\sim 30 Au$, it could perhaps represent
the last remnant of a donor (such as in scenarios 1 and 2), in which case the angular momentum is primordial
and is not included in the discussions above. On the other hand, if the anomalous timing is due to one or
more smaller mass bodies in eccentric orbits, then it would be consistent with the low mass of the planets 
(and their inability to successfully clear the environment of smaller bodies) and the short lifetime of
the system.

\section{Speculations About Other Possible Systems}
\label{Others}

I have concentrated in the previous section on the scenario in which a solar metallicity disk expands and
forms planets. While many scenarios can be analysed in a unified form by simply varying the global
mass and angular momentum budgets (e.g. Phinney \& Hansen 1993), compositional variations are not
so easily attempted. It is not sufficient to simply vary Z, the metal content, in converting gas
to solid bodies. The contribution of heavy elements to the opacity and thus the thermal evolution
of the disk is critical, especially at lower temperatures where the molecular opacities dominate and
which are the regions that especially concern us. Unfortunately, few opacity tables exist for
the more exotic mixtures and we must content ourselves with a few more general speculations.

\begin{figure}
\plotone{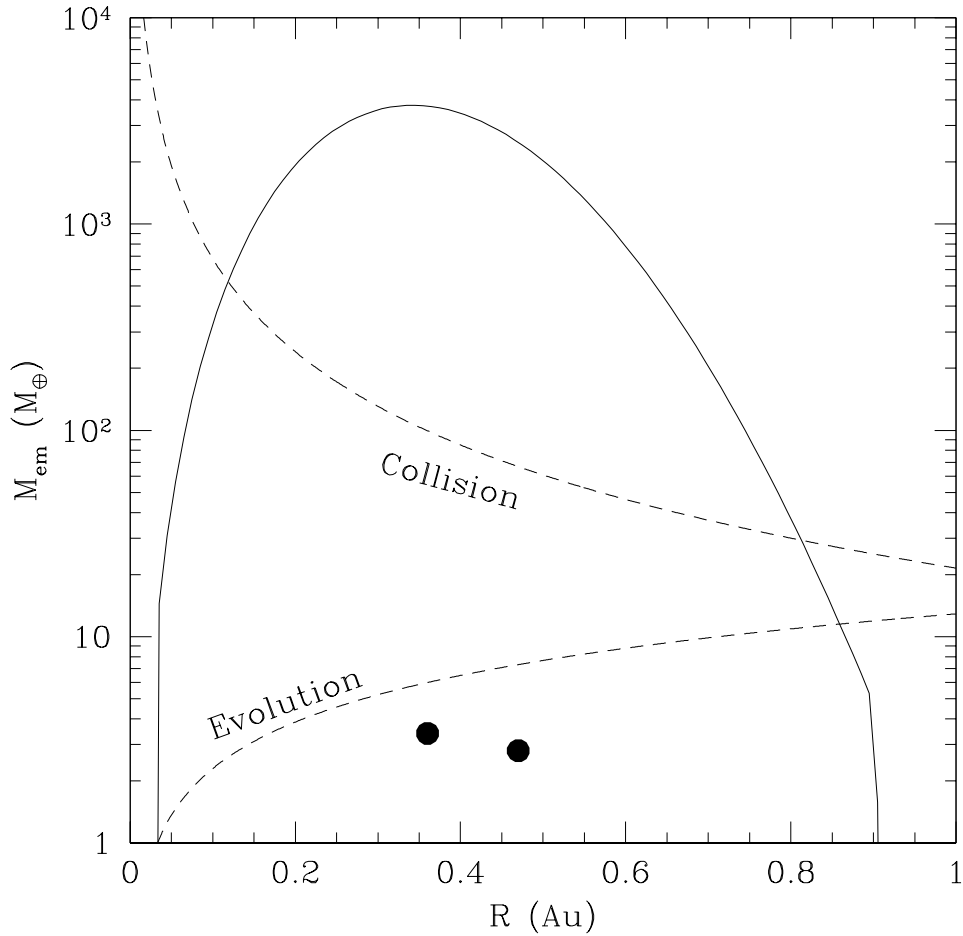}
\caption{ Unlike in the case shown in Figure~\ref{Embryo}, here we find an embarrassment of riches. The fact that
the solid line lies above the curve labelled `Collision' means that bodies are unlikely to build up to this
size by simple agglomeration. Rather, once they cross the upper dashed line they are likely to eject
smaller bodies from the system, rather than accrete them. The pulsar planets are shown at the bottom to
illustrate the difference in the mass scale expected.
\label{Embryo2}}
\end{figure}

This is unfortunate, because many of the most interesting events fall into this category. Perhaps
the most likely candidates are planets formed in the disks resulting from the merger of two
white dwarfs. This is a source population we expect to find and of particular interest in the
case of type Ia supernovae. Even if the central object does not collapse to form a pulsar, we
might expect planets around massive white dwarfs (Livio, Pringle \& Saffer 1992). How might the 
planetary systems we anticipate differ from those discussed above? The first effect worth noting
is the larger expected opacity, which will maintain the disk at a higher temperature. On the
other hand, Carbon has a higher ionization potential than Hydrogen and so a disk composed primarily
of Carbon and Oxygen will recombine at higher temperatures, shutting off the viscosity earlier.
Finally, nearly all of the gaseous disk is potentially available to be converted into planets.
To get a general idea of the expected system, I repeated the calculations above with a toy model where each opacity was
simply increased by a factor 50 ($=1/Z_{\odot}$) and quiescence was assumed to set in below
$10^4$~K. Figure~\ref{Embryo2} shows the equivalent of Figure~\ref{Embryo} in this case, i.e. the
expected size and semi-major axis of bodies expected from a disk of heavy elements.  The fact that
we expect significant mass ejection also suggests that the ejecting bodies will migrate inwards (Murray et al 1998).
Thus, we expect any planets formed in such a disk to be $\sim 30-300 M_{\oplus}$ and located within
$a<0.2$~Au.

Another possibility involves a disk of even heavier elements surrounding a neutron star. This
could arise as in the recent `Fallback disk' model of anomalous X-ray pulsars (Chatterjee, Hernquist
\& Narayan 2000) or, more appropriate to our conference, a remnant disk of neutron-rich material
resulting from the merger of two neutron stars. Such a disk is susceptible to explosion due to the
heat input from radioactive decay (Li \& Paczynski 1998). However, given the deep potential well
of the merger remnant, material inside a radius $R_{bound} \sim \epsilon^{-1} G M/c^2$ (where
$\epsilon$ is the fraction of the nuclear rest mass energy released in the radioactive decay) will
remain bound. For $\epsilon \sim 10^{-3}$, this is $\sim 4 \times 10^8$~cm, yielding a significant
disk with the concomitant possibility of planets. 
The thermal balance of this disk will be complicated
not only by the unusual opacities, but also the additional heat source resulting from the radioactivity.
Planets formed from such material may themselves be directly observable due to heating from radioactivity.
While estimates are obviously wildly uncertain, order of magnitude numbers are
\begin{equation}
L \sim 0.5 L_{\odot} \frac{\epsilon}{10^{-3}} \frac{M_p}{10 M_{\oplus}} \left( \frac{\tau}{10^6 years} \right)^{-1}
\end{equation}
where $\epsilon$ is the fraction of rest mass energy released in radioactivity on some characteristic timescale
$\tau$. Effective temperatures in this case are $\sim 40 000$~K, assuming a radius of $10^9$~cm. The observability
will die away on the timescale $\tau$ as the heat capacity of such objects is small. 
This signature is likely to be quite unique; a body with an {\em extremely} strange spectrum consisting entirely
of heavy elements, in close orbit about a dark
compact object. Note that, if such planets do form, they do so only once the disk has expanded
to radii $> 10^{11}$~cm (the Roche limit for a rocky body around a 3~$M_{\odot}$ compact object).

\section{Conclusion}

The models presented above are obviously quite simple, but it is encouraging that the physically
most well-motivated models give numbers that are in good accord with the observations. However, this
agreement is threatened by the appearance of further timing residuals in the data. If this turns out
to be due to lower mass objects in eccentric orbits then it will represent a striking confirmation
of the picture presented here. However, if it is due to a large object at Neptunian distances, then
it could  invalidate the preceding discussion.

Extrapolating our models to the systems of planets that could result from compact object mergers, we
are limited by our lack of knowledge of the microphysics of gas of such unusual composition. Nevertheless,
we find that quite massive planets in close orbits are possible. These could be found orbiting massive
white dwarfs, neutron stars or black holes.

Support for this work was provided by NASA through Hubble Fellowship grant 
\#HF-01120.01-99A,
from the Space Telescope Science Institute, which is operated by the
 Association of Universities
for Research in Astronomy, Inc., under NASA contract NAS5-26555.


\end{document}